\documentclass[12pt]{article}
\usepackage{rotate}
\usepackage{epsfig}

\begin{document}
\def\oppropto{\mathop{\propto}}
\def\operarrow{\mathop{\longrightarrow}}
\def\opsimeq{\mathop{\simeq}}
\def\opeq{\mathop{=}}
\def \sinc{ \mbox{sinc}}
\newcommand \be  {\begin{equation}}
\newcommand \bea {\begin{eqnarray} \nonumber }
\newcommand \ee  {\end{equation}}
\newcommand \eea {\end{eqnarray}}

\title{Anomalous relaxation in complex systems: from stretched to compressed exponentials}

\author{Jean-Philippe Bouchaud$^{1,2}$}
\maketitle

\small{
$^1$ Service de Physique de l'{\'E}tat Condens{\'e},
Orme des Merisiers -- CEA Saclay, 91191 Gif sur Yvette Cedex, France.
\\
$^2$ Science \& Finance, Capital Fund Management, 6 Bd
Haussmann, 75009 Paris, France.}

\begin{abstract}
We attempt to give a bird's eye view of the physical mechanisms leading to anomalous relaxation, and the relation of this phenomenon with anomalous diffusion and transport. Whereas in some cases these two notions are indeed deeply related, this needs not to be the case. We review several models for stretched exponential relaxation (diffusion in traps, broad distribution of relaxation times, two-step relaxation) and insist on the physical interpretation to be given to the stretching exponent $\beta$. We then discuss compressed exponentials which have been recently observed in a variety of systems, from soft glassy materials to granular packs. We describe a model where slow rearrangement events occur randomly in space and create long ranged elastics strains, leading to $\beta=3/2$. 
\end{abstract}

\section{Introduction}

Normal diffusion usually refers to plain Brownian motion, and describes situations where the mean-square displacement of a particle (or the variance of the (log)-price of a financial asset) grows linearly with time. This property is in general only valid when time is sufficiently large compared to a microscopic correlation time. The diffusion law $R^2 \propto Dt$ is indeed tantamount to a complete decorrelation between successive displacements.
\textit{Anomalous diffusion}, on the other hand, describes all other cases, where the normal diffusion law fails to describe the data, at least over some time
interval $[t_<,t_>]$ where the logarithmic slope $2\nu(t) \equiv \partial \ln R^2/\partial \ln t$ is observed to be either smaller than one (subdiffusion) or larger than one (superdiffusion) \cite{PhysRep}.  Of course, the effect is only interesting if the time interval $[t_<,t_>]$ is wide enough to delimit a genuinely anomalous regime, and not a trivial crossover. One mechanism leading to anomalous diffusion is the existence of persistent (or anti-persistent) correlations up to time interval $t_>$. Positive correlations of displacements lead to superdiffusion, whereas negative correlations lead to subdiffusion. For example, diffusion of particles in hydrodynamical flows often lead to superdiffusion since particles are coherently convected along streamlines over some length scale \cite{PhysRep}. Conversely, particles in a potential field tends to be trapped in minima, where they execute numerous back and forth oscillations, and therefore anti-correlated motion, before being able to escape \cite{PhysRep}. Another important mechanism generating superdiffusion is the existence of  \textit{jumps} with a broad distribution of possible sizes, leading to ``L\'evy flights'' \cite{Levy}. 

Normal diffusion is also characterized by a Gaussian diffusion profile, i.e., the distribution of the displacement $\vec R$ between $t_0$ and $t_0+t$ is given
by:
\be
P(\vec R,t) = \frac{1}{(\sqrt{4\pi Dt})^d} \exp\left[  -\frac{R^2}{4Dt}\right] 
\ee
The relaxation of a density fluctuation of wavevector $\vec q$, created at time $t_0$, is easily computed to be \textit{exponential}:\footnote{Here we imagine that particles do not interact, and that there is no difference between the individual and collective diffusion constant $D$. More generally, there is a difference between the relaxation of tracer particles (of small concentration) and of the density of interacting host particles.}
\be\label{defC}
C(\vec q,t) =\frac{1}{N}\left\langle \sum_j \exp\left[i \vec q \cdot (\vec r_j(t_0+t)-\vec r_j(t_0))\right]  \right\rangle = \exp (-D q^2 t),
\ee    
with relaxation time $1/Dq^2$ equal to the diffusion time over the scale $1/q$. Therefore, normal diffusion is associated to what is usually considered to be ``normal'' -- i.e. exponential -- relaxation. The aim of this short review is to discuss possible forms of anomalous relaxation in complex systems, 
in particular the relation between anomalous diffusion and non exponential relaxation. 

\section{Stretched exponential relaxations}

Stretched exponential relaxation refers to cases where the correlation function decays as $\ln C(t) \propto - t^\beta$ with $\beta < 1$. A simple model where this
behaviour is exactly observed is the case where subdiffusion is induced by an anti-persistent Gaussian process (also called \textit{fractional Brownian motion} \cite{Mandel}), in which case the diffusion profile is given by:
\be
P(\vec R,t) = \frac{1}{(\sqrt{4\pi D_\beta t^\beta})^d} \exp\left[  -\frac{R^2}{4D_\beta t^\beta}\right] , \quad \beta < 1.
\ee
The density relaxation is then trivially given by $C(\vec q,t)=\exp (-D_\beta q^2 t^\beta)$. However, it is hard to think of a physical system for which this model applies directly. A more generic model for subdiffusion is anomalous trapping, for which, as we discuss now, the correlation indeed decays
as a stretched exponential but only at small times.

\subsection{Trapping induced subdiffusion}

We therefore consider the trap model where a particle hops on a regular lattice in $d$ dimensions; each node of the lattice is an energy valley of random depth $E$ which traps the particle for a certain random time, before an activated event occurs and allows the particle to jump to a neighbouring site. One should distinguish the \textit{annealed} case where the depth $E$ is drawn anew at every jump of the particle from the \textit{quenched} case where the depth $E$ of each site is drawn once and for all and does not evolve with time. However, in $d > 2$, the particle typically visits each trap a finite number of times and the difference between the two models is immaterial at long times. We will therefore focus on the first case, which is much easier to solve analytically, and return to the quenched case in $d=1$ at the end of this section.  

Conditioned to being in a trap of depth $E$ and at temperature $T$, the exit time $\tau$ is distributed according to:
\be
\Psi_E(\tau)= \Gamma_0 \ e^{-\beta E} \
 e^{ -\left(\Gamma_0 \ e^{-\beta E} \right)\tau},
\ee
where $\Gamma_0$ is a microscopic frequency and $\beta \equiv 1/T$ (we set $k_B=1$ throughout the paper).
If $\rho(E)$ denotes the \textit{a priori} probability distribution of the depth $E$ of the traps,
 the particle performs a random walk among traps with a distribution of trapping times 
\be
\Psi(\tau)= \int_0^{\infty} dE \ \rho(E) \ \Psi_E(\tau)
\label{trapdistr}
\ee
We will consider below the case where $\rho(E)=1/T_g \exp(-E/T_g)$ (exponential model), for which a true dynamical phase transition appears: diffusion is asymptotically normal for $T > T_g$, but subdiffusion set in for $T < T_g$, with an anomalous diffusion exponent $\nu$ (defined in the introduction) given by $\nu=T/2T_g < 1/2$ . The exponential model is such that the upper limit $t_>$ of the time domain where subdiffusion is observed is infinite. Other cases could be considered as well, such as a Gaussian distribution of depths, but in that case $t_>$  would be finite and a crossover to normal diffusion would eventually set in. However, the following conclusions will still hold in an intermediate time domain, which can become very large when $T \to 0$ (see the detailed discussion in \cite{Monthus}).

Interestingly, the subdiffusive phase $T < T_g$ of the exponential model is also \textit{non stationary}, in the sense that all dynamical properties depend on the age of the system \cite{Dean,Monthus,GBAC}. More precisely, if the initial ($t_0=0$) position of the particle is chosen randomly,  the two-time correlation function
\be C(\vec q, t+t_w,t_w) = \left\langle  e^{i \vec q \cdot \big( \vec r (t+t_w)-
\vec r (t_w) \big)} \right\rangle  \ee 
strongly depends on $t_w$ for $T < T_g$, but becomes time translation invariant and given by Eq. (\ref{defC}) for $T > T_g$, 
at least when $\Gamma_0 t_w \gg 1$. The \textit{aging} properties of the low temperature phase are strongly reminiscent of the phenomenology of 
glasses and spin-glasses below their glass transition \cite{Review}, hence justifying the notation $T_g$ for the critical temperature of the model.

The exact computation of $C(\vec q, t+t_w,t_w)$ was presented in \cite{Monthus}. After a series of manipulations, one obtains the following result 
for the Laplace transform ${\tilde C}(\vec q, \lambda ,t_w) =
\int_0^{\infty} dt \ e^{- \lambda t}  \ C(\vec q, t+t_w,t_w)$:
\be
{\tilde C}(\vec q, \lambda ,t_w) = {\tilde \Pi}( \lambda ,t_w) \
\left[  {1-\zeta_d(\vec q)}
 \over {1-{\tilde \Psi}(\lambda)\zeta_d(\vec q)}  \right]
+{1 \over \lambda} { {1-{\tilde \Psi}(\lambda)} \over {(\zeta_d(\vec q))^{-1}
-{\tilde \Psi}(\lambda)}}
\label{Correlw}
\ee
where we have introduced:
\be
\zeta_d(\vec q)= {1 \over d} \sum_{\kappa=1}^d \cos(q_{\kappa}a)
\ee
and the following Laplace transforms:
\be
{\tilde\Psi}(\lambda)=\int_0^{\infty} d\tau \ e^{- \lambda \tau} \ \Psi(\tau); \quad {\tilde \Pi}( \lambda ,t_w) =
\int_0^{\infty} dt \ e^{- \lambda t} \ \Pi( t+t_w,t_w),
\ee
where $\Pi(t+t_w,t_w)$ is the probability not to have jumped between $t_w$ and $t_w+t$. The above formulas are general and do not depend on the
specific choice of $\rho(E)$. In the case of an exponential density of trap depths, and when $\mu=T/T_g < 1$, the correlation function $\Pi$
is known and its Laplace transform reads \cite{Dean,Monthus}: 
\be
{\tilde \Pi}( \lambda ,t_w) \simeq  {1 \over t_w} \int_0^{\infty} du \
 e^{ \displaystyle - (\lambda t_w) u} \ \  { {\sin \pi \mu} \over {\pi }}
\int_{ { u } \over {1+u}}^{1} dv 
\left(1-v \right)^{\mu -1} v^{- \mu} 
\label{PiexpL}
\ee
with the asymptotic behaviors
\be
{\tilde \Pi}( \lambda ,t_w) \opsimeq_{\lambda t_w \ll 1}
{1 \over {\Gamma(1+\mu)}} {{(\lambda t_w)^{\mu}} \over {\lambda}}
\ee
\be
{\tilde \Pi}( \lambda ,t_w) \opsimeq_{\lambda t_w \gg 1}
{1 \over {\lambda}} \bigg[1-{1 \over {\Gamma(\mu)}} {1 \over {
(\lambda t_w)^{1-\mu}}} \bigg].
\ee
Interestingly, the correlation function $\Pi$ depends on the ratio $t/t_w$ when $\mu < 1$: the characteristic time needed to leave a trap is therefore proportional to the age 
of the system. For $\mu > 1$, on the other hand, one would find that $\Pi$ only depends on $\Gamma_0 t$ and not on $t_w$ -- the problem is \textit{time translation invariant} in that case, as usual for equilibrium dynamics.   

Coming back to Eq. (\ref{Correlw}), it is convenient to introduce the subdiffusion time $t_q$ defined as: 
\be
 (\Gamma_0 t_q)^{\mu}= {2d \over (qa)^2}  
\ee 
corresponding to the typical time needed by the particle to spread over a region of size $1/q$. There are therefore three time scales in (\ref{Correlw}) : $\lambda^{-1}, t_q,t_w$. We are interested in the region where all three are much larger than the
microscopic time scale $\Gamma_0^{-1}$, where the results become universal in the sense that they only depend on the exponential form of $\rho(E)$ at 
large $E$, and not on the details of $\rho(E)$ for finite $E$. Still we have to distinguish various time regimes in (\ref{Correlw}). Let us first consider \textit{young}
systems, for which $t_w \ll t_q,t$. In this case, we find that at small times $t \ll t_q$, the correlation decays as a stretched exponential with exponent $\mu < 1$:\footnote{We define in the following $\sinc x=\sin x/x$.}
\be
\ln C(\vec q, t+t_w,t_w) \simeq -{\sinc [\pi \mu]\over { \Gamma(1+\mu)}}
\left({t \over t_q} \right)^{\mu}
\qquad \hbox{for} \ t_w \ll t \ll t_q,
\ee 
whereas at large times $t \gg t_w$, the correlation decays as a power-law (much slower than a stretched exponential):
\be
C(\vec q, t+t_w,t_w) \simeq \Gamma(1+\mu)
\left({t \over t_q}\right)^{-\mu}
\qquad \hbox{for} \ t_w \ll t_q \ll t 
\ee
In the aging regime, on the other hand, we find at small times:
\be
\ln C(\vec q, t+t_w,t_w) \simeq   
 -    {\sinc [\pi (1-\mu)]}  
\left({t \over t_w}\right)^{1-\mu} \qquad \hbox{for} \ t_q \ll t \ll t_w , 
\ee 
\be
\ln C(\vec q, t+t_w,t_w) \simeq - {\sinc [\pi \mu] \over {\Gamma(\mu)}} 
{t \over {t_q^{\mu} t_w^{1-\mu}}}
\qquad \hbox{for} \   t \ll t_w  \ \ \hbox{and}\  t \ll  t_q   \ 
\ee
and finally, at long times:   
\be
C(\vec q, t+t_w,t_w) \simeq   {\sinc [\pi \mu]} 
\left({t \over t_w}\right)^{-\mu}   
\qquad \hbox{for} \  t_q \ll t_w \ll t   \ 
\ee

There are four interesting points to notice:
\begin{itemize}
\item  For $qa$ very large, such that $t_q \ll t,t_w$, we find the same asymptotic behaviors
 as for $\Pi(t+t_w,t_w )$. Physically, this means that as soon as the particle has jumped once, the
rapidly oscillating correlation function averages to zero. Hence, only the particles
{\it which have not yet moved} contribute to the correlation.
\item  There are two regimes where the correlation function behaves similarly to a
stretched exponential, but \textit{only at small times}, when $t_q \ll t \ll t_w$ or $t_w \ll t \ll
t_q$.  The exponent $\beta$ of this stretched exponential is however different in
both cases: it is equal to $\beta=\mu$ when $t_w \ll t \ll t_q$, and equal to
$\beta=1-\mu$ in the other case.
\item In the regime  $t \ll t_w, t_q$, we find an interesting sub-aging behaviour, where $C(\vec q, t+t_w,t_w)$ is a function of $t/t_w^{1-\mu}$, 
see the discussions in \cite{Review,Maass,Bertin1d}.
\item The frequency dependent susceptibility defined by:
\be
\chi(\vec q,\omega,t_w) = 1 + i\omega \int_0^\infty dt \  e^{i \omega t} 
C(\vec q,t_w+t,t_w) \equiv 1 + i \omega \ \tilde C(\vec q, -i \omega ,t_w),\label{chi}
\ee
behaves
{\it in a Cole-Cole fashion} for young systems: 
\be 
\chi(\vec q,\omega,t_w=0) \simeq {1 \over 1+ (-i\omega t_q)^\mu}
\ee
for $\omega \ll \Gamma_0$, $qa \ll 1$. In the aging regime $\omega t_w \gg 1$, the
behaviour of $\chi$ is given by: 
\be\label{1surf}
\chi(\vec q,\omega,t_w) \simeq { 1\over \Gamma(\mu) (-i \omega t_w)^{1-\mu} }
\ee
A similar expression was obtained in the context of spin-glasses in \cite{Dean}, and discussed further in the context of experimental data in \cite{Review}. 
Note that the noise spectrum $S(\omega)$ of the system is related to the susceptibility through a fluctuation-dissipation relation $S(\omega) \propto \chi''(\omega)/\omega$.
\end{itemize}

Two further remarks on this spatial trap model: a) the marginal case $T=T_g$, $\mu=1$ is quite interesting, since it exhibits exact dynamical ultrametricity \cite{Bertinultra} and weakly aging $1/f$ noise, see Eq. (\ref{1surf}) above and \cite{Dean,Review}; b) the quenched case in $d=1$ reveals a number of subtle peculiarities, discussed in \cite{Bertin1d}. In particular, the diffusion exponent $\nu$ is now given by $\nu=\mu/(1+\mu)$. Using the results of  \cite{Bertin1d}, one finds that in the case of {\it young} systems, the result for the correlation function reads:
\be
\ln C(\vec q, t+t_w,t_w) \sim - \left(\frac{t}{t_{q}}\right)^{\frac{2\mu}{1+\mu}}; \qquad C(\vec q, t+t_w,t_w) \sim \left(\frac{t_q}{t}\right)^\mu,
\ee 
for $t \ll t_q$ and $t \gg t_q$, respectively, with now $t_q$ given by: $qa (\Gamma_0 t_q)^{\nu}=1$. The above shape is different from the annealed case, but the initial decay is still of the stretched exponential type. The asymptotic behaviour of $C(\vec q, t+t_w,t_w)$ in the aging regime can also be inferred from the
results of \cite{BertinF}. We also refer to refs. \cite{traps} for more work on the trap model and some application to glassy dynamics. 

\subsection{Two-step relaxation}

In the above model, there is no intra-node dynamics: traps are assumed to have no spatial extension. This is clearly unrealistic and the short-time, high-q
behaviour of the correlation function should be sensitive to the intra-trap dynamics. As a simple model for this, we consider  harmonic potential wells 
again organized at the node of a regular lattice. Each particle oscillates at the
bottom of the well, and with a small probability per unit time proportional to $\Lambda=\Gamma_0 \exp(-E/T)$ 
the particle hops to its nearest neighbour site, where it lands at
random with the equilibrium probability inside the harmonic well. We neglect for the moment the fluctuations of $E$, which is 
justified when $T \gg T_g$.
The position of a particle at time $t$ will be written as:
\be
\vec r(t) = \vec R(t) + \vec \delta(t),
\ee
where $\vec R(t)$ labels the lattice site to which the particle `belongs' at time $t$, and 
$\vec \delta(t)$ is the position of the particle within the well, the center of the well being 
defined by $\vec \delta(t)=0$. The probability for the particle not jumping site between $t_0$ and 
$t_0 + t$ is 
\be
\Pi(t) = \exp(-\Lambda t).
\ee
If the particle has not jumped, then $\delta(t_0+t) - \delta(t_0)$ is a Gaussian variable of variance 
given by:
\be
\langle (\delta(t_0) - \delta(t_0+t))^2 \rangle = 2\Delta_0^2 [1 - \exp(-\gamma t)],
\ee
where $\Delta_0$ is the width of the explored region of the well, and $\gamma$ the inverse relaxation time
in the well. If on the other hand at least one jump has occurred between $t_0$ and $t_0 + t$, one has:
\be
\langle (\delta(t_0) - \delta(t_0+t))^2 \rangle = 2\Delta_0^2.
\ee
Our aim is again to compute the correlation function:
\be
C(\vec q,t)=\langle e^{i \vec q \cdot (\vec r(t_0) - \vec r(t_0+t))} \rangle.
\ee
This quantity can be easily computed by separating the no jump situations from the situations where at least 
one jump has taken place. One finds:
\be
C(\vec q,t) = \Pi(t) e^{-q^2 \Delta_0^2 [1 - \exp(-\gamma t)]} + (1 - \Pi(t)) 
e^{-q^2 \Delta_0^2} \langle e^{i \vec q \cdot (\vec R(t_0) - \vec R(t_0+t))} \rangle_J,
\ee
where the subscript $J$ indicates that at least one jump has taken place. Clearly, this last average is related to
the unconditional average by:
\be
\langle e^{i \vec q \cdot (\vec R(t_0) - \vec R(t_0+t))} \rangle = \Pi(t) +  (1-\Pi(t)) 
\langle e^{i \vec q \cdot (\vec R(t_0) - \vec R(t_0+t))} \rangle_J.
\ee
Finally, the unconditional average is a classic result of random walk theory on lattices:
\be
\langle e^{i \vec q \cdot (\vec R(t_0) - \vec R(t_0+t))} \rangle = \exp \left[\Lambda t (\zeta_d(\vec q)-1)\right].
\ee
In the limit $q a \ll 1$, the final
result reads (see \cite{Berthier} for similar calculations):
\be
C(\vec q,t) = \Pi(t) \left[e^{-q^2 \Delta_0^2 [1 - \exp(-\gamma t)]}-e^{-q^2 \Delta_0^2}\right]+
e^{-q^2 \Delta_0^2} e^{-\Lambda (qa)^2 t}.
\ee
Note that $C(\vec q,t=0)=C(\vec q=0,t)=1$, as it should.  Suppose $\Lambda t \ll 1$ (i.e. the probability of a jump is very small, $\Pi(t) \approx 1$) and  
$\gamma t \gg 1$, corresponding to full equilibration within the initial well. In this limit, one then finds:
\be
C(\vec q,t) \approx f [1 + q^2 \Delta_0^2 \exp(-\gamma t)],
\ee
i.e. an exponential convergence towards a plateau value $f = e^{-q^2 \Delta_0^2}$. If on the other hand 
$\Lambda t \ll 1$, many jumps have occurred ($\Pi(t) \approx 0$) and:
\be 
C(\vec q,t) \approx f e^{-\Lambda (qa)^2 t}.
\ee
The parameter $f$ measures the relative weight of the two relaxation processes (intra-trap and inter-trap); and has the following interpretation: if one plots 
the above relaxation function in a log-lin representation, one observes two-step relaxation, with a {\it quasi-plateau}
of the correlation function, of height $f$. The fast, intra-trap process does lead to a complete decorrelation -- on the contrary, 
for time scales much smaller than $\gamma^{-1}$, the system appears to be non ergodic (it keeps a non zero correlation
with its initial state); the strength of this non ergodicity is measured by $f$. The quantity $f$ is called the non-ergodicity, or 
Edwards-Anderson parameter, in the context of glassy dynamics. The above simple model also allows to discuss in very simple terms the non-Gaussian parameter $\alpha$ often discussed in this context.  A purely Gaussian probability density is such that $\ln C(\vec q,t) \propto q^2$. The 
kurtosis $\alpha$ is in fact related to the coefficient of the $q^4$ term of the expansion of 
$\ln C(\vec q,t)$ in the vicinity of $\vec q = 0$. One finds that $\alpha(t) \sim [1-\Pi(t)]/t^2$ for small $t$
and $\alpha(t) \sim \Pi(t)/t^2$ at large $t$. In the naive model above, $1-\Pi(t)= \Lambda t$ at small $t$, leading to
a diverging kurtosis at small times. This is however an artefact of the instant `jump' description of the way 
a particle exits from a trap. A more realistic model should account for the fact that the time needed to 
exit a trap cannot be infinitely small, leading to a much smaller value of $1-\Pi(t)$ at small $t$ and therefore 
a vanishing $\alpha(t)$ in that limit. Therefore, we expect $\alpha(t)$ to peak around the typical time needed
to exit a trap, before decaying back to zero at larger times, as often seen in experimental or numerical data (see e.g. \cite{NonG}).

When the fluctuations of the depths $E$ are reinstalled, in particular in the glassy phase $T < T_g$, one finds that the above short time behaviour of $C(\vec q,t)$ is unaffected, while the long time behaviour is still given by the expression of the previous section, up to a factor $f$ which describes the initial fast 
fall-off of the correlation function. 

\subsection{Superposition of relaxation times and stretched exponentials}

In the models considered above, the stretching of the relaxation is associated to (sub-)diffusive entities. This needs not be
the case; in some cases there is relaxation but no transport. Suppose for example that the physics is governed by two-level
systems, rotors or dipoles which do not wander in space at all, but feel a random field which makes the local relaxation time $\tau$
random as well.  More precisely, suppose that there is a whole spectrum of independent, exponential 
relaxation processes, with a density of relaxation times $\rho(\tau)$, and that each elementary relaxation process contributes
equally to the correlation function, so that:
\be\label{Ct}
C(t) = \int_0^\infty \rho(\tau) e^{-t/\tau} d\tau.
\ee
Suppose further that the resulting correlation function $C(t)$ can be well fitted by a stretched exponential:
\be
C(t) \approx \exp - (\gamma t)^\beta,
\ee
what can we infer about the shape of the  density of relaxation times $\rho(\tau)$?
After a very small time, the correlation function has dropped from its initial value of $1$ by an amount 
$\Delta C \sim \int_0^t \rho(\tau) d\tau$:  all relaxation times larger than $t$ have not had time to relax yet, whereas all relaxation times smaller than
$t$ have already relaxed. Equating this result with the initial decay of the stretched exponential, one finds:
\be
\int_0^t \rho(\tau) d\tau \approx (\gamma t)^\beta \longrightarrow \rho(\tau) \propto_{\tau \to 0} 
\gamma^\beta \tau^{\beta-1},
\ee
showing an (integrable) divergence of the density of small relaxation times. 

Therefore, although stretched exponential relaxation is usually associated with \textit{slow} relaxation, the above result implies a {\it 
profusion of short time scales} in the system. However, since the long time behaviour of 
the relaxation is much slower than an exponential, there must also be abundant long time scales. How abundant can
be guessed by a saddle point calculation. Suppose that $\rho(\tau)$ decays itself, for {\it large} $\tau$ as
$\exp(-B \tau^{\beta'})$. This implies, for large $t$, a decay of $C(t)$ given by:
\be\label{stret}
\ln C(t) \simeq - \frac{B^{1-\beta}}{\beta^\beta (1-\beta)^{1-\beta}} t^\beta \qquad \beta \equiv \frac{\beta'}{1+\beta'},
\ee
up to sub-leading power-law corrections. 
For example, a significantly stretched relaxation $\beta=0.5$ corresponds to $\beta'=1$, i.e., an exponential 
decay of long relaxation times \cite{langer}. Therefore, only a very limited density of very large relaxation times is enough
to transform a pure exponential relaxation into a stretched one. The pure exponential case corresponds formally to
$\beta'=\infty$, i.e. no long times at all.

More generally, any non trivial distribution of relaxation times leads to a relaxation function that is faster than
exponential on short times and slower than exponential on large times, in the following sense: define the average 
relaxation time as $\langle \tau \rangle= \int_0^\infty C(t) dt = \int_0^\infty \tau \rho(\tau) d\tau$; the 
reference exponential relaxation is taken to decay over this average time. Then,
the initial slope of the relaxation is given by:
\be
- \frac{dC}{dt} = \int_0^\infty \frac{1}{\tau} \rho(\tau) d\tau = \langle \frac1\tau \rangle 
\geq \frac{1}{\langle \tau \rangle},\qquad ({\mbox{provided}}\quad \langle \frac1\tau \rangle<+\infty)
\ee
where the equality is reached only if there is a single relaxation time scale, $\rho(\tau)=\delta(\tau-\gamma^{-1})$.
Conversely, for large time scales, one can easily show that the relaxation can only be slower than $\exp(-t/\langle \tau \rangle)$. 

An important case, which superficially corresponds to the trap model discussed above, is:
\be
\rho(\tau)= \frac{\mu \tau_0^\mu}{\tau^{1+\mu}}  \Theta(\tau-\tau_0),
\ee
i.e.  a power-law density of relaxation times.The cut-off for small $\tau$ (that could be chosen to be less abrupt) and the condition $\mu > 0$ 
are needed for the distribution to be normalisable. Because of the cut-off, $\langle \frac1\tau \rangle < + \infty$; the initial slope of the $C(t)$ is
 then finite and the short time decay is regular, at variance with the stretched exponential case described above. However, the long time
decay is in that case much slower than stretched exponential:
\be
C(t) = \int_{\tau_0}^\infty  \frac{\mu \tau_0^\mu}{\tau^{1+\mu}} e^{-t/\tau} d\tau \simeq_{t \to \infty}
\Gamma(\mu) \left(\frac{\tau_0}{t}\right)^\mu.
\ee
The above result is valid for all values of $\mu$, in particular when $\mu < 1$, which corresponds to an infinite average relaxation time. 
At this stage, the reader might feel completely nonplussed: why all the {\it aging properties} that we found within the trap model when 
$\mu < 1$ seem to have disappeared? The answer is that the two models are in fact fundamentally different. In the trap model, particles 
progressively discover their environment and, as they explore space further, they fall in deeper and deeper traps which they never encountered before. 
When $T < T_g$, the deepest trap encountered at time $t_w$ is so much deeper than all the others that it dominates the dynamics. In the second
model, on the other hand, there is a relaxing entity for each and every trap, and therefore the whole distribution of relaxation time is probed right away. There
is no aging whatsoever in that case. 

It should be noted that a stretched exponential fit to non exponential relaxation is often an acceptable fit of empirical (or numerical) data. For example, 
when $\tau \propto \exp(E/T)$ with a Gaussian distribution of barrier heights, the relaxation obtained from Eq. (\ref{Ct}) can still be approximatively 
fitted by a stretched exponential, with, as a rough rule of thumb \cite{Monthus}:
\be\label{betaT2}
\beta \approx \frac{T}{\sqrt{T^2 + E_0^2}}
\ee
where $E_0$ is the width of barrier distribution. When a stretched exponential fit is attempted, it is important to specify whether it is supposed to be faithful
to the initial decay of the relaxation $t \leq \gamma^{-1}$, and hence to probe the profusion of high frequencies in the system, or if it is rather the long time ($t \gg \gamma^{-1}$) aspect of the phenomenon which is targeted, which is often the regime where exact analytical results are available but where the correlation is so small $-\ln C(t) \gg 1$ that it nearly impossible to measure. To illustrate this point, is interesting to discuss the problem of survival of a Brownian particle in a random environment, where infinitely deep traps are randomly scattered in a $d$ dimensional space. For this problem, a rigorous result, due to Donsker and Varadhan \cite{DV}, asserts that the total survival probability, averaged over the starting point of the Brownian particle, decays asymptotically as a stretched exponential, with $\beta=d/(d+2)$. This result can be simply understood using the 
above relation, Eq. (\ref{stret}), between $\beta'$ and $\beta$: in the model at hand, long times are associated with exceptionaly large regions free of traps
where the particle can survive much longer than on average.These occur with probability $\ln p \propto -L^d$, and the associated survival time is $\tau(L)\sim L^2/D$, from which one immediately concludes that $\beta'=d/2$. 
Although this result is rigorous, its domain of validity is, in $d \geq 2$, so far into the asymptotic regime that it is of little interest, at least for numerical 
or experimental purposes, because the number of surviving particles in this regime is exponentially small. For intermediate times, one finds that the survival probability can be fitted by a stretched exponential, but with an effective value of $\beta  \neq d/(d+2)$.  We should also mention that this survival model is in fact very subtle: although the average survival probability is given by the above Donsker-Varadhan result, the survival of a given particle with fixed initial positiion decays as $-\ln C(t) \sim t/\ln t$! We refer to the insightful papers of Ben Arous et al. \cite{GBA2,GBA} for a comprehensive account of this result.

\section{Models of compressed exponentials}

Again, a trivial model leading to both superdiffusion and compressed exponential relaxation (i.e. $\beta > 1$) is when the noise driving the particle is a long-ranged persistent Gaussian process, in which case the diffusion profile is given by:
\be
P(\vec R,t) = \frac{1}{(\sqrt{4\pi D_\beta t^\beta})^d} \exp \left[ -\frac{R^2}{4D_\beta t^\beta}\right] , \quad \beta > 1.
\ee
and therefore $C(\vec q,t)=\exp (-D_\beta q^2 t^\beta)$, which is faster than exponential when $\beta > 1$. As we noticed for anti-persistent Brownian motion, 
physical situations where this model applies directly are scarce. In fact, natural models for superdiffusion do {\it not} lead to compressed exponentials, as we now discuss. We then turn to the necessary ingredients needed to explain why compressed exponential relaxations have recently been observed in a variety of soft glassy materials -- where one would have rather expected, because of their glassiness, stretched exponentials! On the other hand, exactly as stretched 
exponential relaxation means super-abundance of short relaxation times, compressed exponential relaxation requires that no fast relaxation channels exist.

\subsection{Superdiffusion and L\'evy flights}

The best known model of superdiffusion is a random walk with a distribution of jump size $\ell$ which decays for large $\ell$ as:
\be
P(\ell) \sim \frac{\mu \ell_{0}^{\mu}}{\ell^{1+\mu}},
\ee
with $\mu < 2$ such that the variance of the distribution diverges.In this case, one knows that for large times, the distribution of the total
displacement of the particle is given by a L\'evy stable law of index $\mu$:\footnote{For simplicity, we restrict here to one dimensional L\'evy
flights.}
\be
P(R,t) \propto \frac{1}{t^\nu} L_\mu(\frac{R}{t^\nu}); \qquad \nu=\frac{1}{\mu},
\ee
where we have been sloppy with prefactors and constants. The crucial point is that, although $L_\mu$ has no simple expression in general (except for
$\mu=2$ where it reduces to the Gaussian and $\mu=1$ where it is the Cauchy distribution), its Fourier transform is extremely simple. In fact, the 
relaxation of a density perturbation is an unspectacular exponential:
\be\label{Levyq}
C(q,t) = \exp (-Dq^\mu t).
\ee
Although diffusion is anomalous, relaxation is normal! The only anomalous feature is the scaling of the wavevector $q$, raised to the power $\mu$ rather than
$q^2$ for normal diffusion. Exactly this behaviour was observed for tracer particles in a system of polydisperse, elongated micelles \cite{Ott}. The very strong contrast between the diffusion constant of small micelles and long micelles was argued to be responsible for the broad distribution of jump sizes, and in turn for the L\'evy flight motion clearly observed experimentally \cite{Ott}.   

Another well studied case of superdiffusion is random advection in long range correlated hydrodynamical flows. Consider the following Langevin equation for a particle in a $d=3$ random flow:
\be
\frac{d\vec x}{dt}= -\vec V(\vec x) + \vec \eta(\vec x,t),
\ee
where $ \vec \eta$ is the usual Langevin noise and $\vec V$ is a divergence free random convection field, with long-ranged correlations:
\be
\left\langle V_i(\vec x) V_j(\vec y) \right\rangle \opsimeq_{ |\vec x - \vec y| \to \infty} {\cal T}_{ij} |\vec x - \vec y|^{-a}, \qquad a < 2,
\ee  
where ${\cal T}$ is a tensor ensuring that $\vec V$ is indeed an incompressible flow. This model, with $a=-2/3$ can be seen as a very rough model of tracer dispersion in a turbulent flow with Kolmogorov scaling-- in order to be more realistic one should allow the flow field to become time dependent, with a correlation that includes a function of $(t-t')/|\vec x - \vec y|^{2/3}$. For the above static turbulence model, one can show that diffusion is anomalous with a diffusion exponent exactly given by $\nu=2/(2+a) > 1/2$ (see \cite{PhysRep}), leading to the Richardson scaling $\nu=3/2$ for $a=-2/3$. The shape of diffusion profile at large
times is not exactly known. However, since the simplest self-consistent re-summation scheme of the perturbation theory for this problem (called Mode-Coupling or Direct Interaction Approximation (DIA) in different contexts \cite{McComb}) gives the exact diffusion exponent $\nu$, one can hope that it also leads to an accurate shape of the 
diffusion profile. More precisely,  the DIA suggests that the Fourier-Laplace transform of $P(\vec R,t)$ reads, in the long wavelength, low frequency limit \cite{PhysRep}:
\be
C(\vec q,\lambda) \approx \frac{1}{\lambda + Dq^{1/\nu}},
\ee 
which is exactly the Fourier-Laplace transform of a L\'evy distribution! The interesting consequence is that diffusion in a long-range correlated hydrodynamical flow is equivalent, at long times and in the scaling regime $R \sim t^\nu$, to a L\'evy flight (provided of course the DIA is trustworthy). But is also means that this model cannot generate compressed exponentials!

\subsection{Rearrangement-induced stress fields in elastic media}

Still, one of the major experimental surprise of the last decade is the discovery of compressed exponential relaxation in a bevy of different soft matter materials, ranging from polystyrene micro-sphere gels, diblock copolymers, laponite, etc. \cite{luca}. These systems have very slow dynamics and rheological properties typical of glasses or jammed granular media, which would have suggested that relaxation functions would be found, as in glasses, to be stretched exponentials.
More precisely, one finds experimentally that the correlation function decays as:
\be
C(\vec q,t) = \exp [-A(qt)^\beta],
\ee  
with $\beta \approx 3/2 > 1$. A notable feature of the above result is the scaling between space and time $q^{-1} \sim t$, characteristic of ballistic effects.
It was suggested early on \cite{lucaearly} that such a result might be related to the elasticity of these materials which mediate the long-ranged stress fields generated when a local rearrangement occurs. In this section, we show how a more detailed model of the deformation induced by these local events can give rise to anomalous $q$ and $t$ behaviour of the structure function \cite{pitard}. 

\subsubsection*{The model}

Following \cite{lucaearly}, we assume that the dominant mechanism is the random appearance
of localized rearrangements. For example, in the system studied in \cite{lucaearly}, the micro-particles forming the gel attract each 
other rather strongly, the gel tends to restructure locally so as to create dense packings of particles. Since the collapsing particles belong to a gel network, 
their motion will induce a certain strain field around them; other particles therefore move and dynamical light 
scattering probes this motion. Similarly, in other soft elastic media, any local rearrangement will induce a strain that propagates into the system.
A rearrangement of particles will result in a force dipole of intensity $P_0$ in a certain direction $\vec n$. When the
dipole is formed at point $\vec r_0$, the elastic strain field $\vec u$ at point $\vec r$ can
be computed assuming a simple central force elasticity:
\be
K \Delta \vec u = - \vec f (\vec r) ,\label{eqn1}
\ee
where $K$ is a compression modulus
and $\vec f (\vec r)$ is a local force dipole of the
form:
\be
\vec f(\vec r)=
f_0 \left[\vec \epsilon \cdot \vec \nabla \delta(\vec r - \vec r_0)\right] 
\vec n,
\ee
where we will assume that $\vec \epsilon = \epsilon \vec n$, i.e, the mean 
displacement of the particles creates a force in the same direction. This dipolar
force can be simply expressed by its Fourier transform:
\be
\vec f(\vec k)= i P_0 \,(\vec k . \vec n)\, \vec n \, \exp(-i\vec k . \vec r_0)
\ee
with $P_0=f_0 \epsilon$.\footnote{Note that a more refined model with shear modulus could be
also be considered, but would only change some numerical factors in the 
following calculations.}
The solution of equation (\ref{eqn1}) is of course:
\be
\vec u(\vec r) = - \frac{P_0}{4\pi K} \frac{(\vec r -\vec r_0)\cdot \vec 
n}{|r-r_0|^3} \vec n.
\ee
The $r^{-2}$ dependence of the strain field has an immediate consequence which 
will be
of importance in the following: if there is a finite density of force dipoles 
randomly
scattered in space, the probability that the stress at a given point has an 
amplitude $u$
decays for large $u$ as $u^{-5/2}$, which has a diverging variance. (This
divergence is
however cut-off if the finite size of the dipoles is taken into account).
This property of the distribution of displacements and stresses will be
responsible for the unusual $q$-dependence of the correlation function.
 
Now, let us assume that the rearrangement events are not instantaneous but take place slowly,
over a certain time scale $\theta$. This will turn out to be crucial in the following. 
A given event  $j$ starts at time $t_j$ and is completed at time $t_j+\theta$;
the dipole intensity $P(t)$ at time $t$ is a certain function of 
$(t-t_j)/\theta$, which can be taken to be approximately linear: $P_j(t)=P_0 
(t-t_j)/\theta$ before saturating at
$P_0$ when the collapse is completed. The dynamics of the individual particles is 
presumably
dominated by viscous friction, therefore we write the following equation of 
motion for the
strain field $\vec u$:
\be
\gamma \frac{\partial \vec u(\vec r,t)}{\partial t} =  
K \Delta \vec u + \sum_j \vec f_{j}(\vec r, t) 
 + \vec \eta(\vec r,t),\label{eqmotion}
 \ee
 where the Fourier transform of the dipolar force is
\be
\vec f_{j}(\vec k, t)=i P_j(t) \,(\vec k . \vec n_j)\, \vec n_j \, \exp(-i\vec 
k. \vec r_j)
\ee
and $\gamma$ is a friction coefficient, and $\vec \eta$ is the thermal random 
force due to the viscous bath, uncorrelated in time and in 
direction. We will finally assume in the following that these events occur randomly in 
space and time,
with a certain rate $\rho$ per unit volume and unit time, and with a random orientation of
the force dipole $\vec n$. The quantity $K/\gamma$ is a diffusion constant that we will call 
$D$. Equation
(\ref{eqmotion}) defines the model that we want to study and from which we 
will compute
the correlation function (or dynamical structure factor) $C(q,t)$, defined as:
\be
C(q,t) = 
\langle \exp\left[i \vec q \cdot \left(\vec u(\vec r,t_0+t) 
- \vec u(\vec r,t_0)\right)\right] \rangle
\ee
where the brackets refer to a spatial average over $\vec r$ or, equivalently, 
over the
random location and time of the rearrangement events. In the following, we 
will neglect
the thermal random force, which would add a Debye-Waller short time diffusive 
contribution to the
dynamical structure factor, and set $\vec \eta = 0$. However
the presence of $\vec \eta$ has an indirect crucial effect since 
the local rearrangements are probably thermally activated.  

\subsubsection*{The slow rearrangement regime}

A first step is to calculate the Fourier transform of the {\it time 
derivative} of
the displacement field $\vec u(\vec r,t)$ created by a single dipole located 
at $\vec r_j$,
in direction $\vec n_j$, that we denote $\vec v(\vec k,t|\vec r_j,\vec 
n_j,t_j)$. One finds:
\bea
\vec v(\vec k,t|\vec r_j,\vec n_j,t_j)& = & -i \exp(-i\vec k\cdot \vec r_j) 
\frac{P_0 \vec n_j}{\theta} \ \frac{\vec n_j \cdot \vec k}{K k^2}
\exp(-Dk^2 t)\\& & \left[\exp(Dk^2 t_j) - \exp(Dk^2 \min(t_j+\theta,t))\right].
\eea
The displacement difference between $t$ and $t+\tau$ can therefore be 
expressed as:
\be
\vec u(\vec r,t_0+t)-\vec u(\vec r,t_0)=
\int_{t_0}^{t_0+t} dt' \sum_{j/t_j < t'} \int \frac{d^3 \vec k}{(2\pi)^3} 
\exp(i\vec k\cdot \vec r) \vec v(\vec k,t_0|\vec r_j,t_j)
\ee
The analysis of this expression reveals that there are several cases
to consider for the relative position of the time $t_j$ when the $j^{th}$ 
event takes place and the other relevant times. We refer to \cite{pitard} for
further details.  In the course of the analysis, one  finds that a new, $q$-dependent time scale $\tau_q$ appears, defined as:
\be
\tau_q \equiv \frac{D\theta}{q v_0} \theta,
\ee
such that, depending on the ratio $t/\tau_q$, 
the dominant contribution to the decay of $C(q,t)$ comes from different 
regions of the $t',t_j$ plane. In the slow regime where the rearrangement time $\theta$ is larger than the 
lag $t$, the final result reads:
\be
C(q,t) = \exp\left[- A'  \rho \theta (D\theta)^{3/2}
\left(\frac{t}{\tau_q}\right)^{3/2}\right] \qquad (t \ll 
\tau_q),\label{result1}
\ee
which has the compressed exponential form found in \cite{lucaearly}, in particular, it 
indeed only depends on $(qt)^{3/2}$. (The numerical value of the prefactor $A'$ can be computed 
exactly to be $1.67996..$.)
This result is however only valid in the short time regime $t \ll \theta$ and $t \ll \tau_q$. Note that the combination $\hat \rho 
\equiv \rho \theta (D\theta)^{3/2}$
is adimensional and represents the average number of events taking place 
within a time
interval $\theta$ and within a diffusion volume $(D\theta)^{3/2}$. 

\subsubsection*{Other regimes}

Other regimes can be analyzed similarly (see \cite{pitard}), and
lead to different scaling behaviours for $C(q,t)$. A summary of the results in terms of 
$\Xi(q,t) \equiv -\log C(q,t)$ is as follows. The two physical cases depend on the
relative position of $\tau_q$ and $\tilde \tau_q=qv_0/D= \theta^2/\tau_q$. 
For $D\theta \ll qv_0$ one finds $\tau_q \ll \theta \ll \tilde \tau_q$ and:
\be 
\Xi \sim (qt)^{3/2} \quad (t \ll \tau_q);\quad \Xi \sim (qt)^{5/4} 
\quad (\tau_q \ll t \ll \tilde \tau_q);\quad
\Xi \sim q^{3/2}t \quad (t \gg \tilde \tau_q).
\ee
Note the appearance of a second compressed exponential regime with $\beta=5/4$.
For $D\theta \gg qv_0$, this $(qt)^{5/4}$ regime is squeezed out and 
the results are simply:
\be 
\Xi \sim (qt)^{3/2} \quad (t \ll \theta);\qquad
\Xi \sim q^{3/2}t \quad (t \gg \theta).
\ee
where we have skipped all prefactors. The late time, purely exponential regime in $q^{3/2}t$ in fact reads:
$\rho v_0^{3/2} q^{3/2} t$, which can be simply understood. The factor $\rho t$
simply counts the average number of events per unit volume between $t_0$ and $t_0+t$, and $q^{3/2}$ reflects
the fact that the distribution of local displacements $u$ decays as $u^{-5/2}$ and has a diverging variance. 
For a distribution with a finite variance, one would obtain the usual $q^2$ dependence.

The mechanism leading to a compressed exponential at short times is the fact that the motion of particles is
not diffusive in this regime, but rather ballistic, due to a local drift of the structure imposed by slow events
taking place elsewhere. Clearly, if $\theta$ was very small (corresponding to instantaneous rearrangements), one would lose the 
compressed exponential behaviour altogether and observe simple exponential decay with an anomalous $q^{3/2}$ scaling, exactly 
as for L\'evy flights (see Eq. (\ref{Levyq})).  The fact that the motion of particles is \textit{coherent} over a rather long time scale $\theta$ is
crucial to observe the small time slow decorrelation $1-C(t) \sim t^{3/2}$. As recently noted \cite{Lech}, this ballistic-like motion might be
a generic property of glassy system, where ``crossing a barrier'' means a slow, collective rearrangement of many particles in an orderly manner.
There might therefore be deep inter-relations between compressed exponential relaxation, cooperative effects and dynamical heterogeneities (for recent 
reviews, see \cite{Ediger,NonG,uslong} and refs. therein).

\section{Conclusion}

We have tried to give an overview of the physical mechanisms leading to \textit{anomalous relaxation}, and the relation of this phenomenon with anomalous diffusion and transport. Whereas in some cases these two notions are indeed deeply related, this needs not to be the case. We have shown in particular that stretched 
exponential relaxation can be found in trap models where subdiffusion occurs, but only as a short time approximation, and in a non stationary (aging) situation. At long times, the relaxation becomes a power-law in time. We have pointed out that stretched exponential relaxation, while being slow at long times, in fact reveals the existence of \textit{fast} relaxation processes at short times. We have emphasized the fact that any distribution of  relaxation times will lead to a relaxation function resembling, at least within some time interval, a stretched exponential; therefore one should be very careful about the physical consequences one infers from a stretched exponential fit of experimental (or numerical) data points, in particular the physical meaning of the stretching exponent $\beta$. 

Models leading to compressed exponentials are much more scarce. In particular, the dynamical structure factor of particles undergoing L\'evy flight motion is a simple exponential in time. Similarly, superdiffusion in random advection flows lead, at least within some approximations, to a simple exponential in time; it would be interesting to know under what conditions this is an exact result. Still, compressed exponentials have now been observed in a variety of systems, from soft glassy materials to granular packs. We have discussed a generic model for compressed exponentials where slow rearrangement events occur randomly in space and create long ranged elastics strains, leading to $\beta=3/2$. 
 
\vskip 1cm
\underline{Acknowledgements}: This review is heavily based on work I have been involved in during the last 15 years or so. I thank in particular G. Ben Arous, E. Bertin, L. Berthier, G. Biroli, M. Cates, L. Cipelletti,  L. Cugliandolo, D. Dean, O. Dauchot, A. Georges, J. Kurchan, F. Lechenault, Ph. Maass, M. M\'ezard, C. Monthus, 
E. Pitard, D. Reichman,  P. Sollich, H. Yoshino \& M. Wyart for many useful discussions on these and related matters. 
I also thank Eric Bertin and Giulio Biroli for carefully reading the manuscript. Finally I want to thank Rainer Klages and Gunter Radons
for giving me the opportunity to write this piece.


\begin{thebibliography}{99}

\bibitem{PhysRep} see e.g. J.-P. Bouchaud, A. Georges, {\it Anomalous diffusion in random media: statistical mechanisms, models and physical applications}, Phys. Rep. {\bf 195} 127 (1990) .
\bibitem{Levy}  {\it L\'evy Flights and Related Topics in Physics}, 
Edts: M. F. Shlesinger, G. M. Zaslavsky, U. Frisch, Lecture Notes in Physics, vol. 450.
\bibitem{Mandel} B. B. Mandelbrot, J. W. Van Ness, {\it Fractional Brownian Motions, Fractional Noises and Applications},  
SIAM Review, {\bf 10}, 422 (1968).
\bibitem{Monthus} C. Monthus, J.-P. Bouchaud, {\it Trap Models and phenomenology of glasses}, J. Phys. A: Math. Gen. {\bf 29} 3847 (1996).
\bibitem{GBAC} G. Ben Arous, J. Cerny, {\it Dynamics of trap models}, math.PR/0603344.
\bibitem{Dean} J.P. Bouchaud, D.S. Dean,  {\it Aging on Parisi's tree},  J. Physique I (France) {\bf 5},  (1995) 265.
\bibitem{Review} J.-P. Bouchaud, L. Cugliandolo, J. Kurchan, M. M\'ezard, {\it Out of equilibrium dynamics in spin-glasses and other glassy systems}; in {\it Spin-glasses and Random Fields}, edited by A.~P. Young (World Scientific, Singapore, 1998).
\bibitem{Bertin1d} E. Bertin, J.P. Bouchaud, {\it Subdiffusion and localization in the one-dimensional trap model}, Phys. Rev. E 67, 026128 (2003).
\bibitem{Maass} B. Rinn, Ph. Maass, J.P. Bouchaud, {\it Hopping in the glass configuration space: Subaging and generalized scaling laws}, Phys. Rev. B 64, 104417 (2001).
\bibitem{Bertinultra} E. Bertin, J.P. Bouchaud, {\it Dynamical ultrametricity in the critical trap model}, J. Phys. A: Math. Gen. 35 (2002) 3039.
\bibitem{BertinF}  E. Bertin, J.P. Bouchaud, {\it Linear and nonlinear response in the aging regime of the one-dimensional trap model}, Phys. Rev. E 67, 065105 (2003).
\bibitem{traps} J.C. Dyre, {\it Master-equation appoach to the glass transition}, Phys. Rev. Lett. {\bf 58} (1987) 792,
Phys. Rev. {\bf B51} (1995) 12 276;  T. Odagaki, {\it Glass Transition Singularities}, Phys. Rev. Lett. {\bf 75} (1995) 3701; 
P. Sollich, F. Lequeux, P. Hebraud, and M.E. Cates, {\it Rheology of soft glassy materials}, Phys. Rev. Lett. {\bf 78}, 2020 (1997)
R. A. Denny, D. R. Reichman, J.P. Bouchaud, {\it Trap Models and Slow Dynamics in Supercooled Liquids}, Phys. Rev. Lett. {\bf 90} 025503 (2003); A. Heuer, B. Doliwa, and A. Saksaengwijit, {\it 
Does the potential energy landscape of a supercooled liquid resemble a collection of traps?}, Phys. Rev. E {\bf 72}, 021503 (2005).
\bibitem{Berthier} L. Berthier,  D. Chandler and J.P. Garrahan. {\it Length scale for the onset of Fickian diffusion in super-
cooled liquids}, Euro. Phys. Lett. {\bf 69}, 320 (2005).
\bibitem{NonG} see e.g.: W. Kob, C. Donati, S. J. Plimpton, P. H. Poole and S. C. Glotzer, {\it Dynamical Heterogeneities in a Supercooled Lennard-Jones Liquid}, Phys. Rev. Lett. {\bf 79} 2827 (1997).
\bibitem{langer} see e.g. J. S. Langer, S. Mukhopadhyay, {\it Anomalous diffusion and stretched exponentials in heterogeneous glass-forming materials},
arXiv:0704.1508 for a very recent contribution on this subject. Older papers include: I. A. Campbell, J. -M. Flesselles, R. Jullien and R. Botet
{\it Nonexponential relaxation in spin glasses and glassy systems}, Phys. Rev. B {\bf 37}, 3825 (1988); J. T. Bendler, M. F. Shlesinger, 
{\it Defect diffusion and a two-fluid model for structural relaxation near the glass-liquid transition}, J. Phys. Chem. {\bf 96}, 3970 (1992).
\bibitem{DV} M. Donsker and S. R. Varadhan, {\it On the number of distinct sites
visited by a random walk} Comm. Pure Appl. Math. {\bf 32}, 721 (1979).
\bibitem{GBA2} G. Ben Arous, S.A. Molchanov, A.F. Ramirez, 
{\it Transition from the annealed to the quenched asymptotics for a random walk on random obstacles}, Ann. Probab. {\bf 33}, 
2149 (2005).
\bibitem{GBA} G. Ben Arous, L.V. Bogachev, S.A. Molchanov, {\it Limit theorems for sums of random exponentials}, Probability Theory and Related Fields,
{\bf 132} 579 (2005).
\bibitem{Ott} A. Ott, J.P. Bouchaud, D. Langevin, W. Urbach, {\it Anomalous diffusion in ''living polymers'': A genuine Levy flight?}, Phys. Rev. Lett. {\bf 65} 2201  (1990); J. Phys. II France {\bf 1} 1465 (1991).
\bibitem{McComb} W. D. McComb, {\it Renormalisation methods}, Oxford University Press, 2004.
\bibitem{luca} L. Cipelletti and L. Ramos, {\it Slow dynamics in glassy soft matter}, J. Phys.: Condens. Matter {\bf 17}, R253 (2005), see also: 
R. Bandyopadhyay, D. Liang, J. L. Harden, R. L. Leheny, {\it Slow dynamics, aging, and glassy rheology in soft and living matter}, e-print cond-mat/0606466.
\bibitem{lucaearly} L. Cipelletti, S. Manley, R.C. Ball, D.A. Weitz, {\it Universal Aging Features in the Restructuring of Fractal Colloidal Gels}, 
Phys. Rev. Lett. 84, 2275 - 2278 (2000). 
\bibitem{pitard} J.-P. Bouchaud, E. Pitard, {\it Anomalous dynamic light scattering in soft glassy gels}, Eur. Phys. J. E {\bf 6} 231 (2001).
\bibitem{Lech} F. Lechenault, O. Dauchot, G. Biroli, J.P. Bouchaud, {\it Experimental evidence of critical scaling and heterogeneous superdiffusion across the Jamming transition}, in preparation.
\bibitem{Ediger} M.~D. Ediger, {\it Spatially heterogeneous dynamics in supercooled liquids}, Annu. Rev. Phys. Chem. {\bf 51}, 99 (2000); R. Richert, {\it Heterogeneous dynamics in liquids: fluctuations in
space and time}, J. Phys.: Condens. Matter {\bf 14}, R703 2002.
\bibitem{uslong} L. Berthier, G. Biroli, J.-P. Bouchaud, W. Kob, K. Miyazaki, and D.~R. Reichman, 
{\it Spontaneous and induced dynamic fluctuations in glass-formers}, cond-mat/0609656 and 0609658, to appear in J. Chem. Phys. 

\end{thebibliography}
\end{document}